\documentclass{article}
\usepackage{fleqn,espcrc2,times}
\usepackage{graphicx}
\usepackage{latexsym}
\usepackage{amssymb}
\DeclareGraphicsExtensions{.eps,.ps}
\title{Short Coherence Length Superconductivity: A Generalization of BCS
  Theory for the Underdoped Cuprates}
  \author{K. Levin, Qijin Chen, and Ioan Kosztin %
  \address{The James Franck Institute, The University of Chicago, 5640
    South Ellis Avenue, Chicago, Illinois 60637}}
\begin{document}

\begin{abstract}
  On the basis of the observed short coherence lengths in the cuprates we
  argue that a BCS--Bose-Einstein condensation (BEC) crossover approach is
  an appropriate starting point for correcting the mean field approach of
  BCS and, thereby, for addressing pseudogap phenomena in these materials.
  Our version of the BCS-BEC approach is based on a particular Greens'
  function decoupling scheme which should be differentiated from others in the
  literature, and which yields (i) the Leggett crossover
  ground state (for all coupling constants $g$, at $T=0$) and (ii) BCS
  theory (for all $ T \leq T_c$ over a range of small $g$).  In this paper
  we provide a simple physical picture of the pseudogap phase above and
  below $T_c$, and review the quantitative and qualitative implications of
  this theory, which, for the most part have been published in a series of
  recent papers. \hfill \textsf{\bf cond-mat/0003133}%
\vspace*{-1mm}
\end{abstract} 

\maketitle

\section{Introduction and Formalism}

While two main alternatives have been suggested for addressing the cuprate
pseudogap (the phase fluctuation approach\cite{Emery} and the ``nodal"
$d$-wave quasi-particle picture\cite{Lee}), this paper presents a third
alternative which is based on the presumption that one should investigate
small excursions from BCS without abandoning it altogether.  This philosophy
rests on the observation that (i) $\xi$ is short so that the mean field
theoretic approach of BCS should not be expected to be correct in detail and
(ii) on Uemura's observation\cite{Uemura} that the cuprate superconductors
belong to a large class of ``exotic" materials. While the possible role of
stripes, spin-charge separation and $d$-wave pairing has received much
attention in the cuprate literature, here we argue that Uemura's famous plot
suggests that one formulate a theory of superconductivity around a more
universal approach without invoking specific materials-dependent features.

In this paper we present an extension of BCS theory, capable of describing
short $\xi$ superconductors, which is intermediate between a BCS and
Bose-Einstein condensation (BEC) picture (the ``BCS--BEC crossover scenario").
At the heart of our approach is the distinction between the fermionic
excitation gap $\Delta$ and superconducting order parameter $\Delta_{sc}$
which distinction holds for $T \neq 0$, both above and below $T_c$. The
difference parameter (or pseudogap energy scale) $\Delta_{pg}^2 = \Delta^2 -
\Delta_{sc}^2$ can be associated with low energy, weakly damped, pair
excitations of finite momentum.

It is essential to note that our generalization of BCS theory is based on
the \textit{particular} ground state of the crossover problem\cite{Leggett}:
the Leggett state ( $ \Psi_{\bf k} = \Pi_{\bf k} ( u_{\bf k} + v_{\bf k}
c^{\dag}_{\bf k} c^{\dag}_{-{\bf k}} ) |0\rangle$), which has a BCS-like
character but should be thought of as applicable to arbitrary values of the
attractive coupling constant $g$.  Here $u_{\bf k}, v_{\bf k}$ are given by
the usual BCS-like expressions in terms of the fermionic dispersion $E_{\bf
  k} $. The variational conditions which ensue from Leggett's calculations
are the same as those written below in Eqs. (1) and (2), for $T=0$.  We have
shown, based on the work of Kadanoff and Martin\cite{Kadanoff} (KM), that
these self consistent equations can be derived from a Green's function
decoupling scheme. We stress here the three important but subtle
observations of KM: (i) to obtain BCS-like theories, it is sufficient to
truncate the system of equations so that only one and two particle
propagators enter and (ii) the correlation functions associated with this KM
scheme -- at the level of the gap equations and thermodynamics -- are
\textit{not} the same correlation functions which enter into the
electrodynamical response.  (iii) The pair susceptibility $\chi$ which
appears in the T-matrix, or pair propagator $ {\cal T} = g / ( 1 + g \chi
)$, is of the form $ \chi = GG_0$; this differs from other T-matrix schemes
(where, frequently, $\chi = GG$), but is required to obtain the BCS-like gap
equations and their thermodynamics\cite{Kadanoff}.  It should also be noted
that this ground state has 100\% condensation for all $g$, unlike that of
the non-ideal Bose gas.

Without any detailed calculations, we can anticipate the form
of the generalized BCS theory, which applies below $T_c$\cite{Kosztin}.
As expected by a natural extension of the Leggett ground state
variational conditions (and, as is consistent with BCS theory) we have
\begin{eqnarray}
  \label{eq:gap1}
  1+g\sum_{\mathbf{k}} \frac{1-
    2f(E_{\mathbf{k}})}{2E_{\mathbf{k}}}\, \varphi^2_{\mathbf{k}}
&=&  0\;,
  \\
 \label{eq:gap2}
 \sum_{\mathbf{k}}\left[1-
    \frac{\epsilon_{\mathbf{k}}}{E_{\mathbf{k}}} +
\frac{2\epsilon_{\mathbf{k}}}{E_{\mathbf{k}}}\,f(E_{\mathbf{k}})\right]
    &=& n
\end{eqnarray}
Here $\epsilon_{\mathbf{k}}$ is the bare fermion dispersion, measured from
the chemical potential $\mu$, $E_{\mathbf{k}} =
\sqrt{\epsilon_{\mathbf{k}}^2 + \Delta^2\varphi_{\mathbf{k}}^2}$ is the
quasi-particle dispersion and $\varphi_{\mathbf{k}}$ represents the symmetry
of the pairing state.  The new physics of the pseudogap phase is embodied in
a third equation which represents the difference of the two energy gap
parameters in terms of the number of excited pairs (with Bose distribution
$b(\Omega_q)$), as
\begin{equation}
 \label{eq:gap3}
   \Delta^2 - \Delta_{sc}^2 =   \Delta_{pg}^2 = a_0 \sum_{\mathbf{q}} b(\Omega_{\mathbf{q}})\;,
\end{equation}
where  the coefficient of proportionality, $a_0$, and the pair excitation
energy, $\Omega_{\mathbf{q}}$, can both be determined from the above
microscopic theory.  These calculations\cite{Kosztin,Maly} indicate that
(at small, but non-zero ($q, \Omega $)) the pair propagator can be
approximated by
\begin{equation}
{\cal T} \approx a_0/[ \Omega - \Omega_q + \mu_{pair} + i\Gamma_q] 
\end{equation}
for the purposes of calculating the self energy $\Sigma$ of the fermions
(which self energy enters into the resulting gap equations and
thermodynamics).  One can interpret this theoretical approach as follows.
Here we go beyond BCS to include self energy $\Sigma$ effects associated
with the incoherent, finite ${\bf q}$ pair excitations.  \textit{Whereas,
  BCS is a mean field treatment of the particles, the present approach
  should be thought of as a mean field treatment of the pairs}. This
represents the next level in a hierarchy of (superconducting) mean field
theories, which hierarchy is similar to that encountered in magnetic
problems. This mean field approximation of the pairs is associated with the
truncation of the equations of motion, so that pair-pair interactions are
not directly present.  Residual inter-boson interactions arise only
indirectly via the self energy of the fermions.  As a consequence, both
above and below $T_c$, the pair dispersion is quadratic $\Omega_q = q^2 /
M_{pair}$, as in a quasi-ideal Bose gas. In this way, as in the KM paper,
the pair correlation function is to be distinguished from that associated
with the collective modes of the order parameter\cite{Kosztin2}.

\section{Results}

The results summarized here represent those deduced from solving the 
coupled equations\cite{Kadanoff} for the pair propagator $\cal T$ and its
single particle analogue ($G$), along with the fermionic number constraint.

\textit{Pseudogap onset:} In our earliest work\cite{Janko} we found that the
temperature $T^*$ at which the Fermi liquid state first breaks down
corresponds to the onset of a splitting of the single peaked (broadened
Fermi liquid-like) electronic spectral function into two peaks separated by
a (pseudo)gap.  This pseudogap state is characterized by the presence of
metastable pairs or ``resonances" which effectively reduce the single
particle density of states.  Near, but above $T_c$, the latter takes the
form of a broadened BCS-like structure.  Moreover, between $T_c$ and $T^*$,
the wave-vector dependence of $\Sigma$ (and of the pseudogap) departs from
that of a strict $\varphi_{\mathbf{k}}$ symmetry\cite{Maly}.  Note, also,
that in the present approach we always have $T^* \geq T_c$.

\textit{Superconducting transition:} As the temperature decreases from
$T^*$, the density of meta-stable pairs with momentum $q$ continues to
increase, until at $T=T_c$, a macroscopic fraction with $q=0$ undergoes Bose
condensation.  At this temperature, the pair propagator or T-matrix diverges
as expected from the Thouless criterion, so that both $\mu_{pair} $ and
$\Gamma_{q=0}$ vanish.  Moreover, near $T_c$, the inverse
lifetime of the $ q\neq 0$ pairs, as well as that associated with fermion
states $\gamma$ becomes very small\cite{Maly}.  This is an important set of
observations.  Once the vicinity of $T_c$ is reached, the interaction
between the long wavelength (soft) bosons and the (gapped) fermions becomes
weak.  This same behavior continues to hold below $T_c$ so that to leading
order, lifetime effects can be dropped in Eqs. (1)-(3).

The dependence of $T_c$ on $g$ which results from Eqs (1)-(3) is highly
non-monotonic, even for the $s$-wave case\cite{Maly}. At low $g$, the curve
for $T_c$ vs $g$ follows the BCS result until $\Delta_{pg}$ becomes
sufficiently large; then $T_c$ decreases, with $g$, thereby reflecting the
difficulty of forming a superconducting state in the presence of a fermionic
gap.  Once $\mu$ becomes negative, $T_c$ then increases with increasing $g$
approaching a (mass) renormalized, but otherwise ideal BEC limit.  For the
$d$-wave case, the situation is even more complex\cite{Chen1}.  As a result
of the extended size of the $d$-wave pairs, the effects of the Pauli
principle repulsion are more extreme, and $T_c$ vanishes well before the
system reaches the $ \mu < 0 $ limit (except in the limit of unphysically
small densities).  In strictly $2d$, we find $T_c$ is always zero.

\textit{Behavior of the superconducting gap equations:} Equations (1)-(3)
were solved to yield the results plotted in Figure 1, for the case of weak,
moderate and large $g$ . Also indicated is a ``cartoon" characterizing the
excitations of the condensate at each value of $g$. Above $T_c$, where the
computations are more complicated we used a simple extrapolation
procedure\cite{Kosztin2} to facilitate the numerics.  As can be seen
from the Figure and from Eq.~(3), as the
number of excited pair states increases, so does the difference between the
fermionic excitation gap and the order parameter.  These figures represent a
natural generalization of BCS theory.

\textit{Thermodynamical consequences; mostly $C_v$:} The pair excitations
shown schematically in Figure 1 necessarily lead to corrections to the
quasi-particle-derived thermodynamics of BCS theory, via new low $T$ power
laws.  Indeed, it would be difficult to imagine how fermionic
quasi-particles could be relevant to the thermodynamics of the BEC limit.
That these excited pair states have a nearly ideal Bose gas character is a
consequence of the BCS-like, Leggett ground state; they arise from a mean
field treatment of the pair propagator.  One might imagine an improved
approach to the strong coupling limit (which includes direct pair-pair
interactions) would lead to results which are more analogous to those of the
non-ideal Bose gas.  It should, however, be noted that the cuprates are in
the clear fermionic regime where $\mu$ is essentially $E_F$, as is
consistent with our calculations\cite{Chen1,Chen2}.  Moreover, for the
$d$-wave case, our analysis shows that the BEC limit is virtually
inaccessible, and concerns about inaccuracies of the KM approach, in this
limit, appear to be less relevant.  What is most important is to properly
capture the physics of the BCS regime, so that small excursions from it can
be considered in a controlled manner. As a result of these pair excitations,
in a quasi-2d system such as the cuprates we find\cite{Chen2a} $ C_v =
\alpha T^2 ~+ \gamma^* T $.  At this time, a low $T$, linear specific heat
contribution is well documented in the cuprates, although it is not known
whether it is intrinsic or extrinsic.  Elsewhere in this
journal\cite{Chen3}, we compare our quantitative calculations for $C_v$ with
experiment.

\begin{figure}
\vspace*{2mm}
\hskip -1.8mm
\mbox{\parbox{2.12in}{
\includegraphics[width=2.12in, clip]{AtDr100-x-6-2}
}}\mbox{\parbox{.9in}{
\includegraphics[height=.9in, clip, bb = 0 85 510 245, angle=-90]{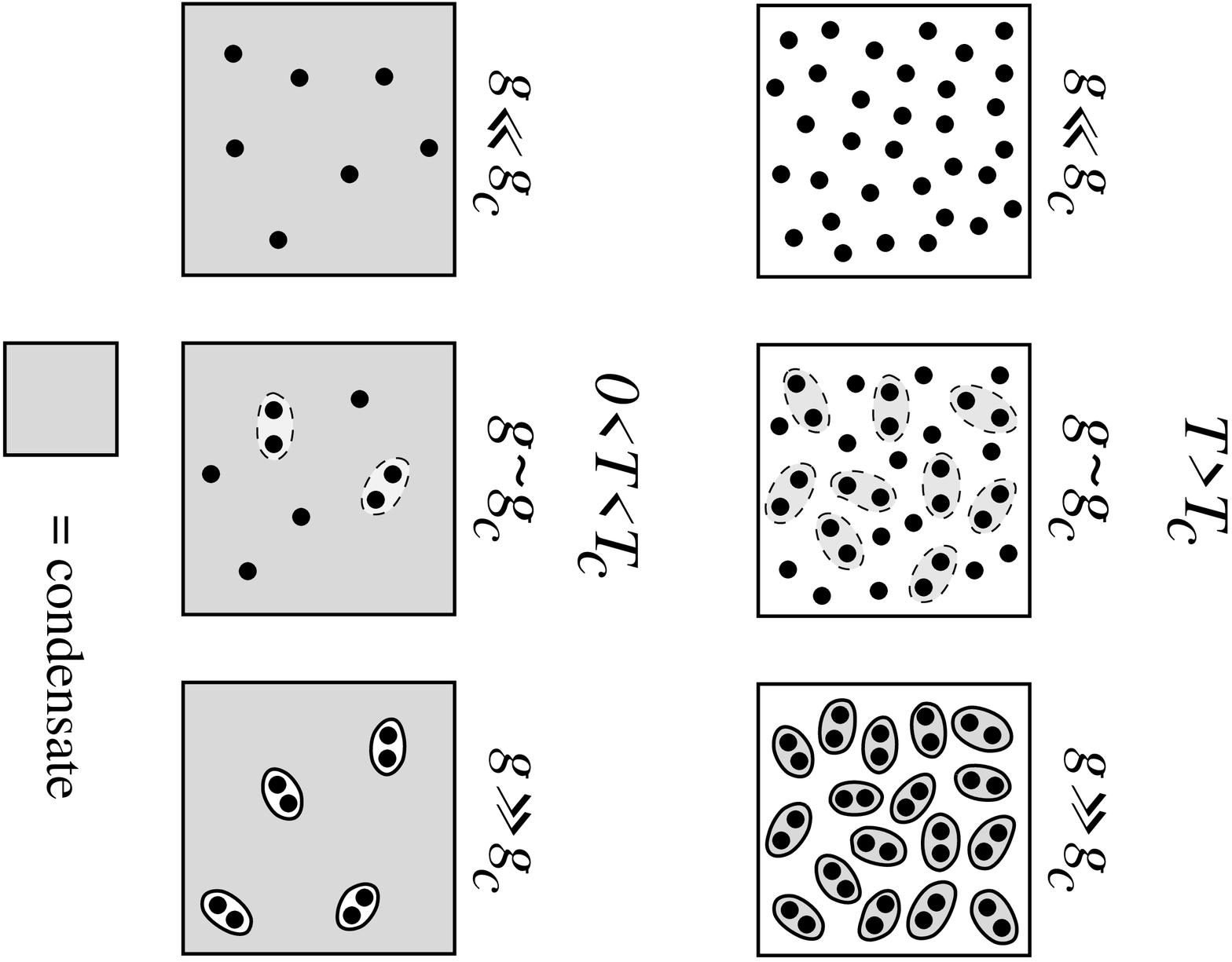}
\vspace*{4.5ex}
}}
\vspace*{-10mm}
\caption{\small Temperature dependence of the gaps.
  Pair excitations become progressively more important with increasing
  coupling strength.} 
\vspace*{-5mm}
\end{figure}

We have not yet completed a full calculation of the thermodynamical
properties such as $C_v$ from above to below $T_c$.  Nevertheless, it can be
anticipated on physical grounds that $ C_v/T$ starts to decrease once $T^*$
is reached and that the overall behavior of that contribution to $C_v$ which
is associated with the fermionic degrees of freedom, is rather similar to
that of a much-broadened BCS theory above $T_c$.  \textit{ The density of
  states is depleted precisely at $E_F$, once the temperature passes below
  $T_c$}.  This long range order-induced depletion, is, thereby, reflected
in thermodynamical properties, such as $C_v$ jumps which become
progressively weaker the stronger the coupling $g$.

\textit{Behavior of the penetration depth:} We find\cite{Chen2,Chen2a}, that
the penetration depth at low $T$, behaves as $\lambda = \lambda_0 + AT + B
T^ {3/2}$.  Here the second and third terms represent respectively the
fermionic quasi-particle and the bosonic contributions.  This additional
$T^{3/2}$ bosonic contribution may be difficult to distinguish from the
``dirty" $d$-wave $ T^2$ term, which was previously invoked in fits to the
data, at the lowest $T$. A quantitative discussion of this term is given
elsewhere in this journal\cite{Chen3}, along with comparisons to the data.
[The hole concentration ($x$) dependence must be also addressed, as will be
summarized below].  It should be stressed that this bosonic term is
responsible for the quasi-universal scaling of the penetration
depth\cite{Chen2}, which we have reported.  Indeed, our calculations
indicate that there are systematic deviations from perfect scaling, and the
general trends for these deviations appear to have been observed by the
Cambridge group\cite{Panagopoulos}.

The present approach should be contrasted with the ``$d$-wave nodal
quasiparticle" picture (in its Fermi liquid rendition\cite{Millis}) where
Landau parameters $F_{1s}$ are thought to be important below $T_c$.  In the
present context, these $F_{1s}$ effects are not as primary as is
incorporating the new (non Fermi-liquid) physics of the pseudogap state.
Ultimately Landau effects can be added here, as elsewhere, for
detailed fits to data.

\textit{Phase Diagram:} In order to incorporate the Mott insulator
constraint, the band-width or alternatively $n/m^*$ must be fit to the
$x$-dependence of the zero temperature penetration depth\cite{Chen2}. In the
absence of any detailed information about the source of the pairing
attraction $g$ (which presumably derives from Coulomb interactions\cite{Liu}
in some direct or indirect form, in the $d$-wave channel), we assume that
$g$ is $x$-independent and fit its ratio to the ``bare bandwidth" via one
adjustable parameter, which is chosen to optimize agreement with the entire
phase diagram\cite{Chen1,Chen2}. With this transcription, one can see that
the ratio of $g$ to the effective Fermi energy must increase as the
insulator is approached, so that the bosonic degrees of freedom become more
evident with underdoping.  This prescription provides a quite good fit to
$T^*, T_c, \Delta(0)$ over the entire range of $x$. It also can be used as a
``tool" for addressing the $x$ and $T$ dependence of a wide collection of
experimental data, including the critical current $I_c$, Knight shift and
NMR relaxation rate\cite{Chen2a}. All of these can be written in terms of
the contributions from ``three fluids": the condensate (via $\Delta_{sc}$),
the fermionic excitations (via $\Delta$), and the pair excitations (via
$\Omega_q$, or $\Delta_{pg}$).

\textit{Neutron scattering and relation to condensation energy:}
Elsewhere\cite{kao}, and in this journal\cite{kao2} we have shown that the
incommensurate and commensurate peaks in the neutron cross section can be
interpreted as reflecting the $d$-wave superconducting gap.  Moreover, above
$T_c$, the pseudogap will lead to a residue of these peaks, albeit
broadened.  Within the present picture the neutron resonance (commensurate
structure) and the behavior of the specific heat with its discontinuity at
and around $T_c$ are both consequences of the same underlying pseudogap
physics, but \textit{not} directly of each other.  Moreover, in contrast to BCS
theory, where one can extrapolate the zero field normal state to estimate
the condensation energy, here we find that because of the non-Fermi liquid
nature of the normal state (and its instability at $T_c$) such
extrapolations are problematical.  This (along with other features, such as
coherent and incoherent contributions to ARPES and tunneling, and general
finite magnetic field effects on the pseudogap) will be discussed in more
detail in future work.

\section{Acknowledgements}
This work was supported by the NSF under awards No. DMR-91-20000
(through STCS) and No. DMR-9808595 (through MRSEC).


\begin{thebibliography}{10}
\small
\bibitem{Emery}
V.J. Emery \& S.A. Kivelson, Nature {\bf 374},  434  (1995).

\bibitem{Lee}
P.A. Lee \& X.-G. Wen, Phys. Rev. Lett. {\bf 78},{\footnotesize }4111(1997).

\bibitem{Uemura}
Y.J. Uemura, Physica C \textbf{282-287}, 194 (1997).

\bibitem{Leggett}
A.J. Leggett, J. Phys. (Paris) {\bf 41}, C7/19 (1980).

\bibitem{Kadanoff}
L.P. Kadanoff \& P.C. Martin, Phys. Rev. {\bf 124}, 670 (1961).

\bibitem{Kosztin}
I. Kosztin, Q.J. Chen, B. Jank\'o, \& K. Levin, Phys. Rev. B {\bf 58},  R5936
  (1998).

\bibitem{Maly}
J. Maly, B. Jank\'o, \& K. Levin, Phys. Rev. B {\bf 59},  1354  (1999);
Physica C \textbf{321}, 113 (1999).

\bibitem{Kosztin2}
I. Kosztin, Q.J. Chen, Y.-J. Kao, \& K. Levin, cond-mat/9906180; Phys. Rev. B
  {\bf 61}, in press.

\bibitem{Janko}
B. Jank\'o, J. Maly, \& K. Levin, Phys. Rev. B {\bf 56}, R11407 (1997).

\bibitem{Chen1}
Q.J. Chen, I. Kosztin, B. Jank\'o, \& K. Levin, Phys. Rev. B {\bf 59},  7083
  (1999).

\bibitem{Chen2}
Q.J. Chen, I. Kosztin, B. Jank\'o, \& K. Levin, Phys. Rev. Lett. {\bf 81},
  4708  (1998).

\bibitem{Chen2a}
Q.J. Chen, I. Kosztin, \& K. Levin, cond-mat/9908362, (submitted to Phys. Rev. Lett.).

\bibitem{Chen3}
Q.J. Chen, I. Kosztin, \& K. Levin, cond-mat/9912314, (also in this M2S-HTSC-VI conference proceedings).

\bibitem{Millis}
A.J. Millis,
S.M. Girvin, L.B. Ioffe, \& A.I. Larkin, 
J. Phys. Chem. Solids
\textbf{59}, 1742(1998). 

\bibitem{Liu}
D.Z. Liu \& K. Levin, Physica C {\bf 275},  81  (1997).

\bibitem{kao}
Y.-J. Kao, Q.M. Si, \& K. Levin, cond-mat/9908302, (submitted to Phys. Rev. Lett.).

\bibitem{kao2}
Y.-J. Kao, Q.M. Si, \& K. Levin, cond-mat/0002317, (also in this M2S-HTSC-VI conference proceedings).

\bibitem{Panagopoulos}
Panagopoulos \textit{et al}, Phys. Rev. B \textbf{60}, 14617 (1999).

\end{thebibliography}
\end{document}